\documentstyle[12pt,epsf]{article}
\newcommand{\be}{\begin{equation}}
\newcommand{\ee}{\end{equation}}

\newcommand{\beqa}{\begin{eqnarray}}
\newcommand{\eeqa}{\end{eqnarray}}
\newcommand{\nn}{\nonumber}

\newcommand{\eqref}[1]{(\ref{#1})}

\def\boxit#1{\vbox{\hrule\hbox{\vrule\kern8pt
\vbox{\hbox{\kern8pt}\hbox{\vbox{#1}}\hbox{\kern8pt}}
\kern8pt\vrule}\hrule}}
\def\mathboxit#1{\vbox{\hrule\hbox{\vrule\kern8pt\vbox{\kern8pt
\hbox{$\displaystyle #1$}\kern8pt}\kern8pt\vrule}\hrule}}

\def\IB{\relax\hbox{$\inbar\kern-.3em{\rm B}$}}
\def\IC{\relax\hbox{$\inbar\kern-.3em{\rm C}$}}
\def\ID{\relax\hbox{$\inbar\kern-.3em{\rm D}$}}
\def\IE{\relax\hbox{$\inbar\kern-.3em{\rm E}$}}
\def\IF{\relax\hbox{$\inbar\kern-.3em{\rm F}$}}
\def\IG{\relax\hbox{$\inbar\kern-.3em{\rm G}$}}
\def\IGa{\relax\hbox{${\rm I}\kern-.18em\Gamma$}}
\def\IH{\relax{\rm I\kern-.18em H}}
\def\IK{\relax{\rm I\kern-.18em K}}
\def\IL{\relax{\rm I\kern-.18em L}}
\def\IP{\relax{\rm I\kern-.18em P}}
\def\IR{\relax{\rm I\kern-.18em R}}
\def\IZ{\relax\ifmmode\mathchoice
{\hbox{\cmss Z\kern-.4em Z}}{\hbox{\cmss Z\kern-.4em Z}}
{\lower.9pt\hbox{\cmsss Z\kern-.4em Z}} {\lower1.2pt\hbox{\cmsss
Z\kern-.4em Z}}\else{\cmss Z\kern-.4em Z}\fi}

\def\II{\relax{\rm I\kern-.18em I}}

\def\CA {{\cal A}}

\def\CG {{\cal G}}

\def\CL {{\cal L}}

\def\CU {{\cal U}}

\begin{document}

\hfill  NRCPS-HE-08-10

\vspace{1cm}
\begin{center}
{\Large ~\\{\it Interaction of Non-Abelian Tensor Gauge Fields
}

}

\vspace{2cm}

{\sl George Savvidy\\
Demokritos National Research Center\\
Institute of Nuclear Physics\\
Ag. Paraskevi, GR-15310 Athens,Greece  \\
}
\end{center}
\vspace{2cm}

\centerline{{\bf Abstract}}

\vspace{12pt}

\noindent
Recently we introduced an extended vector bundle X on which
non-Abelian tensor gauge fields realize a connection.
Our aim here is to introduce interaction of non-Abelian tensor gauge fields
with fermions and bosons. We have found that there exist two series
of gauge invariant forms describing this interaction. The linear sum of
these forms comprises the general gauge invariant Lagrangian.
Studying the corresponding Euler-Lagrange equations we found
that a particular linear combination of these forms
exhibits enhanced symmetry which guarantees the conservation of the
corresponding high-rank currents. A possible mechanism of symmetry breaking
and mass generation of tensor gauge bosons is suggested.


\newpage

\pagestyle{plain}


\section{\it Introduction}

It is appealing to extend Yang-Mills theory \cite{yang,chern} so that it will define the
interaction of  fields which carry not only non-commutative internal charges, but
also arbitrary  large spins. This extension
will induce the interaction of matter fields mediated by charged
gauge quanta carrying spin larger than one \cite{Savvidy:2005fi}.
In our recent approach these gauge fields are defined as
rank-$(s+1)$ tensors
\cite{Savvidy:2005fi,Savvidy:2005zm,Savvidy:2005ki,Barrett:2007nn}
$$
A^{a}_{\mu\lambda_1 ... \lambda_{s}}(x)
$$
and are totally symmetric with respect to the
indices $  \lambda_1 ... \lambda_{s}  $.  A priory the tensor fields
have no symmetries with
respect to the first index  $\mu$.
The index $s$ runs from zero to infinity.
The first member of this family of the tensor gauge bosons is the Yang-Mills
vector boson $A^{a}_{\mu}$. This is an essential departure from the
previous considerations, in which the higher-rank tensors were totally symmetric
\cite{fierz,fierzpauli,minkowski,yukawa1,wigner,schwinger,Weinberg:1964cn,chang,singh,fronsdal}.

The extended non-Abelian gauge transformation of the tensor gauge fields
\cite{Savvidy:2005fi,Savvidy:2005zm,Savvidy:2005ki} is defined
by the equation (\ref{polygauge}) and
comprises a closed algebraic structure, because
the commutator of two transformations can be expressed in the form
$$
[~\delta_{\eta},\delta_{\xi}]~A_{\mu\lambda_1\lambda_2 ...\lambda_s} ~=~
-i g~ \delta_{\zeta} A_{\mu\lambda_1\lambda_2 ...\lambda_s},
$$
where the gauge parameters $\{\zeta\}$ are given by the matrix commutators (\ref{gaugealgebra}).
This allows to define generalized field strength tensors (\ref{fieldstrengthparticular})
$
G^{a}_{\mu\nu,\lambda_{1}...\lambda_{s}}
$
which are {\it transforming homogeneously} (\ref{fieldstrenghparticulartransformation})
with respect to the extended gauge transformations (\ref{polygauge}).
The field strength tensors $G^{a}_{\mu\nu ,\lambda_{1}....\lambda_{s}}$
are used to construct two infinite series of gauge invariant quadratic forms
$$
{{\cal L}}_{s}~~, ~~{{\cal L}}^{'}_{s}~~~~~~~~~~ s=2,3,...
$$
Each term of these
infinite series is separately gauge invariant with respect to the extended gauge
transformations (\ref{polygauge}). These forms
contain quadratic kinetic terms  and terms  describing
nonlinear interaction of Yang-Mills type.
In order to make all tensor gauge fields dynamical one should add
all these forms together. Thus the gauge invariant
Lagrangian describing dynamical tensor gauge bosons of all ranks
has the form \cite{Savvidy:2005fi,Savvidy:2005zm,Savvidy:2005ki}
\be\label{generalgaugedensity}
{{\cal L}} = \sum^{\infty}_{s=1}~ g_{s} {{\cal L}}_{s}~+
~ \sum^{\infty}_{s=2}g^{'}_{s} {{\cal L}}^{'}_{s}~,
\ee
where ${{\cal L}}_{1} \equiv {{\cal L}}_{YM}$ is the Yang-Mills Lagrangian.

It is important that:  i) {\it the Lagrangian does not
contain higher derivatives of tensor gauge fields
ii) all interactions take place
through the three- and four-particle exchanges with dimensionless
coupling constant g  iii) the complete Lagrangian contains all higher-rank
tensor gauge fields and should not be truncated iv) the invariance with respect to
the extended gauge transformations does not fix the coupling constants $g_{s}$
and $g^{'}_{s}$}.

The coupling constants $g_{s}$ and $g^{'}_{s}$ remain arbitrary because
every term of the sum is separately gauge invariant and the extended gauge symmetry
alone does not fix them. There is a freedom to vary these
constants without breaking the extended gauge symmetry (\ref{polygauge}). The main
point here is that one can achieve the enhancement of the extended gauge
symmetry properly tuning the coupling constants $g_{s}$ and $g^{'}_{s}$.
Indeed, considering a linear sum of two gauge invariant forms in (\ref{generalgaugedensity})
$$
g_{2}{{\cal L}}_{2}+ g^{'}_{2}{{\cal L}}^{'}_{2},
$$
which describe the rank-2 tensor gauge field $A^{a}_{\mu\lambda}$, we found
\cite{Savvidy:2005fi,Savvidy:2005ki} that for
$$
g^{'}_{2} = g_{2}
$$
the sum $ {{\cal L}}_{2}+  {{\cal L}}^{'}_{2} $
exhibits invariance with respect to
a bigger gauge group (\ref{fullgroupofextendedtransformation}).
In addition to the extended gauge
group (\ref{polygauge}), which we had initially, we get a bigger gauge group
with double number of gauge parameters \cite{Savvidy:2005fi,Savvidy:2005ki,Barrett:2007nn}.
Considering the second pair of quadratic forms in (\ref{generalgaugedensity})
$$
g_{3}{{\cal L}}_{3}+ g^{'}_{3}{{\cal L}}^{'}_{3}
$$
which describe the rank-3 tensor gauge field $A^{a}_{\mu\lambda\rho}$, we found
in \cite{Savvidy:2006at} that for
$$
 g^{'}_{3} = {4\over 3} g_{3}
$$
the system also has an enhanced gauge symmetry (\ref{fullgroupofextendedtransformation}).
The explicit description of these symmetries
together with the corresponding field equations is given in \cite{Savvidy:2006at}.

Our aim now is to extend this construction to a system of interacting
tensor gauge fields with higher-spin fermion and boson  fields.
The fermions are defined as Rarita-Schwinger spinor-tensors \cite{rarita,singh1,fronsdal1}
$$
\psi^{\alpha}_{\lambda_1 ... \lambda_{s}}(x)
$$
with mixed transformation properties of Dirac four-component wave
function (the index $\alpha$ denotes the Dirac index) and
are totally symmetric tensors of the rank $s$
over the indices $\lambda_1 ... \lambda_{s}$.
All fields of the $\{ \psi \}$ family
are isotopic multiplets belonging to the
same representation $\sigma$ of the compact Lie group G
(the corresponding indices are suppressed).
The bosons are defined as totally symmetric Fierz-Pauli rank-s tensors
\cite{fierzpauli}
$$
\phi_{\lambda_1 ... \lambda_{s}}(x)
$$
all belonging  to the
same representation $\tau$ of the compact Lie group G.

We shall demonstrate that the gauge invariant Lagrangian for fermions
and bosons also contains two infinite series of quadratic forms
and the general Lagrangian is a linear sum of these forms. For fermions it takes the form
\be\label{fermiforms}
{{\cal L}}^{F} =\sum^{\infty}_{s=0}~ f_{s } ~{{\cal L}}_{s +1/2}~+
\sum^{\infty}_{s=1}~ f^{'}_{ s }~ {{\cal L}}^{'}_{s+1/2 }
\ee
and for bosons it is
\be
{{\cal L}}^{B} =\sum^{\infty}_{s=0}~ b_{s } ~{{\cal L}}^{B}_{s}~+
\sum^{\infty}_{s=1}~ b^{'}_{ s}~ {{\cal L}}^{' B}_{s }.
\ee
Again it is important to notice that the invariance with respect to
the extended gauge transformations does not fix the coupling constants
$f_{s},~f^{'}_{ s }$ and $b_{s},~b^{'}_{s}$.
The coupling constants $f_{s},~f^{'}_{ s}$ and $b_{s},~b^{'}_{s}$
remain arbitrary. Every term of the sum is separately gauge invariant
and the extended gauge symmetry alone does not define them.
{\it The basic principle which we shall pursue
in our construction will be to fix these coupling constants demanding
realization of enhanced symmetries and unitarity of the theory}\footnote{
For that one should study the spectrum of the theory and its dependence
on these coupling constants. For some particular values of coupling constants
the linear sum of these forms  may exhibit symmetries with respect to a
bigger gauge group $\CG \supset G $.}.

In the second section we shall outline the transformation properties
of non-Abelian tensor gauge fields, the definition of
the corresponding field stress tensors, the general expression for the invariant Lagrangian
and its enhanced symmetries
\cite{Savvidy:2005fi,Savvidy:2005zm,Savvidy:2005ki}.
In the third, forth and fifths sections we shall incorporate into the theory fermions of
half-integer spins.  We shall construct two infinite series of gauge
invariant forms (\ref{fermiforms}).  The invariant Lagrangian is a linear sum of all these forms
and describes interaction of non-Abelian tensor gauge
fields with half-integer spin fermions. At
special values of the coupling constants it shows up enhanced symmetries
and therefore defines conserved tensor currents.
In the sixth, seventh and eighth  sections the above construction will be
extended to include integer-spin  boson fields and a possible
symmetry breaking  mechanism to generate masses of tensor gauge bosons is
suggested.

\section{\it Non-Abelian Tensor Gauge Fields}

The gauge fields are defined as rank-$(s+1)$ tensors
\cite{Savvidy:2005fi}
$$
A^{a}_{\mu\lambda_1 ... \lambda_{s}}(x),~~~~~s=0,1,2,...
$$
and are totally symmetric with respect to the
indices $  \lambda_1 ... \lambda_{s}  $.  A priory the tensor fields
have no symmetries with
respect to the first index  $\mu$. The index $a$ numerates the generators $L^a$
of the Lie algebra $\breve{g}$ of a {\it compact}\footnote{The algebra $\breve{g}$
possesses an orthogonal
basis in which the structure constants $f^{abc}$ are totally antisymmetric.}
Lie group G.

One can think of these tensor fields as appearing in the
expansion of the extended gauge field $\CA_{\mu}(x,e)$ over the unite  vector
$e_{\lambda}$
\cite{Savvidy:2005ki}:
\be\label{gaugefield}
{\cal A}_{\mu}(x,e)=\sum_{s=0}^{\infty} {1\over s!} ~A^{a}_{\mu\lambda_{1}...
\lambda_{s}}(x)~L^{a}e_{\lambda_{1}}...e_{\lambda_{s}}.
\ee
The gauge field $A^{a}_{\mu\lambda_1 ... \lambda_{s}}$ carries
indices $a,\lambda_1, ..., \lambda_{s}$ labeling the generators of {\it extended current
algebra $\CG$ associated with compact Lie group G.} It has infinite many generators
$L^{a}_{\lambda_1 ... \lambda_{s}} = L^a e_{\lambda_1}...e_{\lambda_s}$ and
the corresponding algebra is given by the commutator \cite{Savvidy:2005ki}
\footnote{See also the alternative Abelian expansions in
\cite{yukawa1,wigner,Bengtsson:2004cd,Edgren:2005gq} and the algebras based
on diffeomorphisms
group in \cite{Bakas:1990xu,Ivanov:1979ny}.}
\be
[L^{a}_{\lambda_1 ... \lambda_{s}}, L^{b}_{\rho_1 ... \rho_{k}}]=if^{abc}
L^{c}_{\lambda_1 ... \lambda_{s}\rho_1 ... \rho_{k}}.
\ee
Because $L^{a}_{\lambda_1 ... \lambda_{s}}$ are
space-time tensors, the full algebra includes the Poincar\'e generators $P_{\mu},~M_{\mu\nu}$.
They act on the space-time components of the above generators as follows:
\beqa\label{extensionofpoincarealgebra}
~&&[P_{\mu},~P_{\nu}]=0,\nn\\
~&&[M_{\mu\nu},~P_{\lambda}] = \eta_{\nu\lambda }~P_{\mu}
- \eta_{\mu\lambda  }~P_{\nu} ,\nn\\
~&&[M_{\mu \nu}, ~ M_{\lambda \rho}] = \eta_{\mu \rho}~M_{\nu \lambda}
-\eta_{\mu \lambda}~M_{\nu \rho} +
\eta_{\nu \lambda}~M_{\mu \rho}  -
\eta_{\nu \rho}~M_{\mu \lambda} ,\nonumber\\
~&&[P_{\mu},~L^{a}_{\lambda_1 ... \lambda_{s}}]=0, \nn\\
~&&[M_{\mu \nu}, ~ L^{a}_{\lambda_1 ... \lambda_{s}}] =
\eta_{\nu \lambda_1}~L^{a}_{\mu \lambda_2... \lambda_{s}}
-\eta_{\mu\lambda_1}~L^{a}_{\nu\lambda_2... \lambda_{s}}
+.....+
\eta_{\nu \lambda_s}~ L^{a}_{\mu \lambda_1... \lambda_{s-1}} -
\eta_{\mu \lambda_s}~L^{a}_{\nu \lambda_1... \lambda_{s-1}} ,\nonumber\\
~&&[L^{a}_{\lambda_1 ... \lambda_{s}}, L^{b}_{\rho_1 ... \rho_{k}}]=if^{abc}
L^{c}_{\lambda_1 ... \lambda_{s}\rho_1 ... \rho_{k}}.
\eeqa
It is an extension of the Poincar\'e algebra by generators which contains
isospin algebra G. In some sense the new vector variable $e_\lambda$ plays  a role
similar to the grassmann variable $\theta$ in supersymmetry algebras \cite{Coleman:1967ad,Haag:1974qh}.

{\it The extended non-Abelian gauge transformations of the
tensor gauge fields are defined
by the following equations } \cite{Savvidy:2005zm}:
\beqa\label{polygauge}
\delta A^{a}_{\mu} &=& ( \delta^{ab}\partial_{\mu}
+g f^{acb}A^{c}_{\mu})\xi^b ,~~~~~\\
\delta A^{a}_{\mu\nu} &=&  ( \delta^{ab}\partial_{\mu}
+  g f^{acb}A^{c}_{\mu})\xi^{b}_{\nu} + g f^{acb}A^{c}_{\mu\nu}\xi^{b},\nonumber\\
\delta A^{a}_{\mu\nu \lambda}& =&  ( \delta^{ab}\partial_{\mu}
+g f^{acb} A^{c}_{\mu})\xi^{b}_{\nu\lambda} +
g f^{acb}(  A^{c}_{\mu  \nu}\xi^{b}_{\lambda } +
A^{c}_{\mu \lambda }\xi^{b}_{ \nu}+
A^{c}_{\mu\nu\lambda}\xi^{b}),\nn\\
.........&.&............................ ,\nn
\eeqa
where $\xi^{a}_{\lambda_1 ... \lambda_{s}}(x)$ are totally symmetric gauge parameters.
These extended gauge transformations
generate a closed algebraic structure. To see that, one should compute the
commutator of two extended gauge transformations $\delta_{\eta}$ and $\delta_{\xi}$
of parameters $\eta$ and $\xi$.
The commutator of two transformations can be expressed in the form \cite{Savvidy:2005zm}
\be\label{gaugecommutator}
[~\delta_{\eta},\delta_{\xi}]~A_{\mu\lambda_1\lambda_2 ...\lambda_s} ~=~
-i g~ \delta_{\zeta} A_{\mu\lambda_1\lambda_2 ...\lambda_s}
\ee
and is again an extended gauge transformation with the gauge parameters
$\{\zeta\}$ which are given by the matrix commutators
\beqa\label{gaugealgebra}
\zeta&=&[\eta,\xi]\\
\zeta_{\lambda_1}&=&[\eta,\xi_{\lambda_1}] +[\eta_{\lambda_1},\xi]\nn\\
\zeta_{\nu\lambda} &=& [\eta,\xi_{\nu\lambda}] +  [\eta_{\nu},\xi_{\lambda}]
+ [\eta_{\lambda},\xi_{\nu}]+[\eta_{\nu\lambda},\xi],\nn\\
......&.&..........................\nn
\eeqa
Each single field $A^{a}_{\mu \lambda_1...\lambda_s}(x),~s=2,3,...$ has no geometrical interpretation,
but all these fields together with $A^a_{\mu}(x)$ have geometrical interpretation in terms of {\it connection on
the extended vector bundle X} \cite{Savvidy:2005ki}. Indeed,
one can define the extended vector bundle X whose structure group
is $\CG$ with group elements
$$
U(\xi)=exp[~i \xi(x,e)~],
$$
where
$$
\xi(x,e)=  \sum_s {1\over s!}~\xi^{a}_{\lambda_1 ... \lambda_{s}}(x) ~~L^{a}e_{\lambda_{1}}...e_{\lambda_{s}}.
$$
Defining the extended gauge transformation of $\CA_{\mu}(x,e)$  in a standard way
\be\label{extendedgaugetransformation}
\CA^{'}_{\mu}(x,e) = U(\xi)  \CA_{\mu}(x,e) U^{-1}(\xi) -{i\over g}
\partial_{\mu}U(\xi) ~U^{-1}(\xi),
\ee
we get the extended vector bundle X on which the gauge field
$\CA^{a}_{\mu}(x,e)$ is a connection \cite{chern}.
The expansion of (\ref{extendedgaugetransformation}) over the vector
$e_\lambda$ reproduces gauge transformation law of the tensor gauge  fields
(\ref{polygauge}). Using the commutator of the covariant derivatives
$
\nabla^{ab}_{\mu} = (\partial_{\mu}-ig \CA_{\mu}(x,e))^{ab}
$
\be
[\nabla_{\mu}, \nabla_{\nu}]^{ab} = g f^{acb} \CG^{c}_{\mu\nu}~,
\ee
we can define the extended field strength tensor
\be\label{fieldstrengthgeneral}
\CG_{\mu\nu}(x,e) = \partial_{\mu} \CA_{\nu}(x,e) - \partial_{\nu} \CA_{\mu}(x,e) -
i g [ \CA_{\mu}(x,e)~\CA_{\nu}(x,e)]
\ee
which transforms homogeneously:
$
\CG^{'}_{\mu\nu}(x,e)) = U(\xi)  \CG_{\mu\nu}(x,e) U^{-1}(\xi).
$
Thus the {\it generalized field strengths  are defined as} \cite{Savvidy:2005zm}
\beqa\label{fieldstrengthparticular}
G^{a}_{\mu\nu} &=&
\partial_{\mu} A^{a}_{\nu} - \partial_{\nu} A^{a}_{\mu} +
g f^{abc}~A^{b}_{\mu}~A^{c}_{\nu},\\
G^{a}_{\mu\nu,\lambda} &=&
\partial_{\mu} A^{a}_{\nu\lambda} - \partial_{\nu} A^{a}_{\mu\lambda} +
g f^{abc}(~A^{b}_{\mu}~A^{c}_{\nu\lambda} + A^{b}_{\mu\lambda}~A^{c}_{\nu} ~),\nn\\
G^{a}_{\mu\nu,\lambda\rho} &=&
\partial_{\mu} A^{a}_{\nu\lambda\rho} - \partial_{\nu} A^{a}_{\mu\lambda\rho} +
g f^{abc}(~A^{b}_{\mu}~A^{c}_{\nu\lambda\rho} +
 A^{b}_{\mu\lambda}~A^{c}_{\nu\rho}+A^{b}_{\mu\rho}~A^{c}_{\nu\lambda}
 + A^{b}_{\mu\lambda\rho}~A^{c}_{\nu} ~),\nn\\
 ......&.&............................................\nn
\eeqa
and transform homogeneously with respect to the extended
gauge transformations (\ref{polygauge}). The field strength tensors are
antisymmetric in their first two indices and are totally symmetric with respect to the
rest of the indices.
The inhomogeneous extended gauge transformation (\ref{polygauge})
induces the homogeneous gauge
transformation of the corresponding field strength
(\ref{fieldstrengthparticular}) of the form \cite{Savvidy:2005zm}
\beqa\label{fieldstrenghparticulartransformation}
\delta G^{a}_{\mu\nu}&=& g f^{abc} G^{b}_{\mu\nu} \xi^c  ,\\
\delta G^{a}_{\mu\nu,\lambda} &=& g f^{abc} (~G^{b}_{\mu\nu,\lambda} \xi^c
+ G^{b}_{\mu\nu} \xi^{c}_{\lambda}~),\nonumber\\
\delta G^{a}_{\mu\nu,\lambda\rho} &=& g f^{abc}
(~G^{b}_{\mu\nu,\lambda\rho} \xi^c
+ G^{b}_{\mu\nu,\lambda} \xi^{c}_{\rho} +
G^{b}_{\mu\nu,\rho} \xi^{c}_{\lambda} +
G^{b}_{\mu\nu} \xi^{c}_{\lambda\rho}~),\nn\\
......&.&..........................\nn
\eeqa
The symmetry properties of the field strength  $G^{a}_{\mu\nu,\lambda_1 ... \lambda_s}$
remain invariant in the course of this transformation.

These tensor gauge fields and the corresponding field strength tensors allow to
construct two series of gauge invariant quadratic forms. The
first series is given by the formula \cite{Savvidy:2005zm}:
\beqa\label{fulllagrangian1}
{{\cal L}}_{s+1}&=&-{1\over 4} ~
G^{a}_{\mu\nu, \lambda_1 ... \lambda_s}~
G^{a}_{\mu\nu, \lambda_{1}...\lambda_{s}} +.......\nonumber\\
&=& -{1\over 4}\sum^{2s}_{i=0}~a^{s}_i ~
G^{a}_{\mu\nu, \lambda_1 ... \lambda_i}~
G^{a}_{\mu\nu, \lambda_{i+1}...\lambda_{2s}}
(\sum_{p's} \eta^{\lambda_{i_1} \lambda_{i_2}} .......
\eta^{\lambda_{i_{2s-1}} \lambda_{i_{2s}}})~,
\eeqa
where the sum $\sum_p$ runs over all nonequal permutations of $i's$, in total $(2s-1)!!$
terms and the numerical coefficient is
$$
a^{s}_i = {s!\over i!(2s-i)!}~.
$$
The second series of gauge invariant quadratic forms is given by the formula
\cite{Savvidy:2005fi,Savvidy:2005ki}:
\beqa\label{secondfulllagrangian}
{{\cal L}}^{'}_{s+1}&=&{1\over 4} ~
G^{a}_{\mu\lambda_1,\lambda_2  ... \lambda_{s+1}}~
G^{a}_{\mu\lambda_2,\lambda_{1} ...\lambda_{s+1}} +.......\nonumber\\
&=& {1\over 8}\sum^{2s+1}_{i=1}~a^{s}_{i-1} ~
G^{a}_{\mu\lambda_1,\lambda_2  ... \lambda_i}~
G^{a}_{\mu\lambda_{i+1},\lambda_{i+2} ...\lambda_{2s+2}}
(\sum^{'}_{p's} \eta^{\lambda_{i_1} \lambda_{i_2}} .......
\eta^{\lambda_{i_{2s+1}} \lambda_{i_{2s+2}}})~,
\eeqa
where the sum $\sum^{'}_p$ runs over all nonequal permutations of $i's$, with exclusion
of the terms which contain $\eta^{\lambda_{1},\lambda_{i+1}}$.

In order to make all tensor gauge fields dynamical one should add
the corresponding kinetic terms. Thus the invariant
Lagrangian describing dynamical tensor gauge bosons of all ranks
has the form
\be\label{fulllagrangian2}
{{\cal L}} = \sum^{\infty}_{s=1}~ g_{s } {{\cal L}}_{s }~+
\sum^{\infty}_{s=2}~ g^{'}_{s } {{\cal L}}^{'}_{s }~,
\ee
where ${{\cal L}}_{1} \equiv {{\cal L}}_{YM}$.

As we already noticed in the Introduction the invariance with respect to
the extended gauge transformations
does not fix the coupling constants $g_{s}$ and $g^{'}_{s}$. Therefore
we can tune these coupling constants demanding maximal possible
symmetry of the sum. We found in
\cite{ Savvidy:2005ki,Savvidy:2006at} that
the coupling constants should be chosen as
$
g^{'}_{2} = g_{2},~g^{'}_{3} = {4 \over 3} g_{3}.
$
The free part of the Lagrangian is invariant with respect to the large gauge
group of transformations with additional gauge parameters $\zeta^{a}_{\mu},
\zeta^{a}_{\mu\nu}$:
\beqa\label{fullgroupofextendedtransformation}
\delta A^{a}_{\mu } &=&\partial_{\mu} \xi^{a},\nn\\
\delta A^{a}_{\mu \lambda} &=& \partial_{\mu} \xi^{a}_{\lambda}+
\partial_{\lambda} \zeta^{a}_{\mu} ,\nn\\
\delta A^{a}_{\mu\nu\lambda} &=& \partial_{\mu} \xi^{a}_{\nu\lambda}
+\partial_{\nu} \zeta^{a}_{\mu\lambda}+
\partial_{\lambda} \zeta^{a}_{\mu\nu},
\eeqa
where the gauge parameters $\zeta^{a}_{\mu\lambda}$ should fulfil the
constraint $\partial_{\rho}\zeta^{a}_{\rho\lambda}-
\partial_{\lambda} \zeta^{a}_{ \rho\rho}=0$ .
The coupling constants $g_2$ and $g_3$ remain arbitrary
and define mixing amplitudes between lower- and higher-rank
tensor gauge bosons.  They have to be
fixed by additional physical requirements imposed on these
amplitudes. We shall return to this problem later.

\section{\it First Series of Gauge Invariant Forms for Fermions}

The fermions are defined as Rarita-Schwinger spinor-tensor fields
\cite{rarita,singh1,fronsdal1}
\be\label{raritaschwingerspinfields}
\psi^{\alpha}_{\lambda_1 ... \lambda_{s}}(x)
\ee
with mixed transformation properties of Dirac four-component wave
function  and are totally symmetric tensors of the rank $s$
over the indices $\lambda_1 ... \lambda_{s}$
(the index $\alpha$ denotes the Dirac index and will be
suppressed in the rest part of the article).
All fields of the $\{ \psi \}$ family
are isotopic multiplets  $\psi^{i}_{\lambda_1 ... \lambda_{s}}(x)$
belonging to the same representation $\sigma^{a}_{ij}$ of the compact Lie group G
(the index $i$ denotes the isotopic index).
One can think of these spinor-tensor fields as appearing in the
expansion of the extended fermion field $\Psi^{i}(x,e)$ over the unit
tangent vector
$e_{\lambda}$ \cite{Savvidy:2005fi,Savvidy:2005ki}
\be\label{fermionfield}
\Psi^{i}(x,e) = \sum^{\infty}_{s=0}~
\psi^{i}_{\lambda_1 ... \lambda_{s}}(x) ~e_{\lambda_1}...e_{\lambda_s}.
\ee
Our intention is to introduce gauge invariant interaction of
fermion fields with non-Abelian tensor gauge fields.
The transformation of the fermions under the extended isotopic group we
shall define by the formula \cite{Savvidy:2005zm}
\beqa
  \Psi^{'}(x,e) &=& \CU(\xi) \Psi(x,e),
\eeqa
where
$$
\CU(\xi) = \exp (i g \xi(x,e)) ,~~\xi(x,e)=   \sum^{\infty}_{s=0}~
\xi^{a}_{\lambda_1 ... \lambda_{s}}(x) ~\sigma^a   e_{\lambda_1}...e_{\lambda_s}
$$
and $\sigma^{a}$ are the matrices of the
representation $\sigma$ of the compact Lie
group G, according to which all $\psi's$ are transforming. In components the transformation of
fermion fields under the extended isotopic group therefore will be \cite{Savvidy:2005zm}
\beqa\label{mattertransformation}
\delta_{\xi}  \psi  &=& i g \sigma^{a}\xi^{a}  \psi ,\nonumber\\
\delta_{\xi}  \psi_{\lambda} &=& i g \sigma^{a}( \xi^{a} ~\psi_{\lambda}   +
\xi^{a}_{\lambda}~ \psi) ,\nonumber\\
\delta_{\xi}  \psi_{\lambda\rho} &=& i g \sigma^{a}( \xi^{a} ~\psi_{\lambda\rho}   +
\xi^{a}_{\lambda}~ \psi_{\rho} + \xi^{a}_{\rho}~
\psi_{\lambda} + \xi_{\lambda\rho} ~ \psi),\\
........&.&.......................,\nn
\eeqa
The covariant derivative of the fermion field is defined as usually:
\be
\nabla_\mu \Psi = i \partial_\mu \Psi +g \CA_{\mu}(x,e) \Psi,
\ee
and  transforms  homogeneously:
\beqa\label{covariantmattertransformation}
\nabla_\mu \Psi \rightarrow \CU ~ \nabla_\mu \Psi ,
\eeqa
where we are using the matrix notation for the gauge fields
$\CA_{\mu} = \sigma^{a} \CA^{a}_{\mu}$. Therefore the gauge invariant
Lagrangian has the following form:
\be\label{generalfermionlagrangian}
\CL^F = \bar{\Psi} \gamma_{\mu} [i \partial_\mu \Psi +g \CA_{\mu}] \Psi.
\ee
Expanding this Lagrangian over the vector variable $e_{\lambda}$ one can get a
series of gauge invariant forms for half-integer fermion fields:
\be\label{firstlagrangian}
\CL^F = \sum^{\infty}_{s=0} f_s \CL_{s +1/2},
\ee
where $f_s$ are coupling constants.
The lower-spin invariant Lagrangian is for the spin-1/2 field:
\be
{{\cal L}}_{1/2} =   \bar{\psi}^i  \gamma_{\mu} (\delta_{ij} i\partial_{\mu} ~+~
g \sigma^{a}_{ij} A^{a}_{\mu} )\psi^j = \bar{\psi}  ( i \not\!\partial +
g \not\!\!A )\psi
\ee
and for the spin-vector field $\psi_{\mu}$ together with the additional
rank-2 spin-tensor $\psi_{\mu\nu}$ the invariant Lagrangian has the form \cite{Savvidy:2005zm}:
\beqa\label{firstfermionlagrangianthreehalf}
{{\cal L}}_{3/2} &=&
\bar{\psi}_{\lambda} \gamma_{\mu} ( i\partial_{\mu} + g A_{\mu} )\psi_{\lambda} +
{1\over 2}\bar{\psi} \gamma_{\mu} (i\partial_{\mu} + g A_{\mu} )\psi_{\lambda\lambda}+
{1\over 2}\bar{\psi}_{\lambda\lambda} \gamma_{\mu}
(i\partial_{\mu} + g A_{\mu} )\psi\nonumber\\
&+& g \bar{\psi}_{\lambda} \gamma_{\mu}  A_{\mu\lambda} \psi
+g \bar{\psi}  \gamma_{\mu} A_{\mu\lambda} \psi_{\lambda}
+{1\over 2}g \bar{\psi} \gamma_{\mu}  A_{\mu\lambda\lambda} \psi ~,
\eeqa
and it is invariant under simultaneous
extended gauge transformations of the fermions (\ref{mattertransformation}) and
tensor gauge fields (\ref{polygauge}):
$$
\delta {{\cal L}}_{3/2} =0.
$$
The currents are given by the variation of the action over the tensor gauge fields:
\beqa\label{firstcurrent}
J^{a}_{\mu} &=& g\{\bar{\psi}_{\lambda}  \sigma^{a} \gamma_{\mu}  \psi_{\lambda}+
{1\over 2}\bar{\psi}_{\lambda\lambda}  \sigma^{a} \gamma_{\mu}  \psi  +
{1\over 2}\bar{\psi}  \sigma^{a} \gamma_{\mu}  \psi_{\lambda\lambda}\},\nn\\
J^{a}_{\mu\nu} &=& g\{ \bar{\psi}  \sigma^{a} \gamma_{\mu}  \psi_{\nu } +
\bar{\psi}_{\nu}  \sigma^{a} \gamma_{\mu}  \psi\},\nn\\
J^{a}_{\mu\lambda\rho} &=& {1\over 2}g
\bar{\psi}  \sigma^{a} \gamma_{\mu}  \psi ~ \eta_{\lambda\rho}.
\eeqa
From extended gauge invariance it follows that they are divergenceless with respect to the
first indices:
\be
 \partial_{\mu} J^{a}_{\mu}= \partial_{\mu}J^{a}_{\mu\nu} =
 \partial_{\mu}J^{a}_{\mu\lambda\rho} =0.
\ee

In the next section we shall see that there exists a second invariant Lagrangian
${{\cal L}}^{'}_{3/2}$ which can be constructed in terms of these spinor-tensor fields
and the total Lagrangian is a
linear sum of these two forms $ f_1 {{\cal L}}_{3/2} + f^{'}_1~ {{\cal L}}^{'}_{3/2} $.
The coupling constants $f_{1}$ and $f^{'}_{1}$ remain arbitrary because
every term of the sum is separately gauge invariant and the extended gauge symmetry
alone does not fix them. There is a freedom to vary these
constants without breaking the extended gauge symmetry. We can expect that
one can achieve the enhancement of the extended gauge
symmetry properly tuning the coupling constants $f_{1}$ and $f^{'}_{1}$.
And indeed, as we shall see, in this way one can achieve the fermion currents
conservation with respect to all their indices.
This property is necessary in order to have consistent interaction
with non-Abelian tensor gauge fields.

\section{\it The Second Series of Invariant Forms for Fermions}

The Lagrangian (\ref{firstfermionlagrangianthreehalf}) is not the
most general Lagrangian which can be constructed in terms
of the above spinor-tensor fields (\ref{raritaschwingerspinfields}).
As we shall see, there exists a second invariant Lagrangian ${{\cal L}}^{'}_{F}$
which can be constructed in terms of spinor-tensor
fields (\ref{raritaschwingerspinfields}),  and the total Lagrangian is a
linear sum of the two Lagrangians: $ f ~{{\cal L}}_F + f^{'}~ {{\cal L}}^{'}_F $.
For the lower-spin case we shall demonstrate that  the total Lagrangian
$f_1 {{\cal L}}_{3/2}+ f^{'}_{1}{{\cal L}}^{'}_{3/2} $ exhibits an enhanced
gauge invariance with specially chosen coefficients $ f^{'}_1$.

First, we shall construct general Lagrangian density for arbitrary
higher-rank spinor-tensor fields which contains two terms:
$ f_s~ {{\cal L}}_{s+1/2} + f^{'}_{s}~ {{\cal L}}^{'}_{s+1/2}~~$, s=1,2,...
Indeed, let us consider the gauge invariant tensor density of the form
\cite{Savvidy:2005fi,Savvidy:2005ki}
\be\label{generalfermiondensity}
{{\cal L}}_{\rho_1\rho_2} =    \bar{\Psi}(x,e)  \gamma_{\rho_1} [ i\partial_{\rho_2} ~+~
g \sigma^{a} \CA^{a}_{\rho_2}(x,e) ]\Psi(x,e) .
\ee
It is gauge invariant tensor density because its variation is equal to zero:
\beqa
\delta {{\cal L}}_{\rho_1\rho_2}(x,e) = i\bar{\Psi}(x,e)\xi(x,e)
\gamma_{\rho_1} [ i\partial_{\rho_2} ~+~
g  \CA_{\rho_2}(x,e) ]\Psi(x,e) +\nn\\
+ \bar{\Psi}(x,e)\gamma_{\rho_1}g (-{1\over g}) [\partial_{\rho_2}\xi(x,e)   -ig
[\CA_{\rho_2}(x,e), \xi(x,e) ]\Psi(x,e) +\nn\\
-i\bar{\Psi}(x,e)\gamma_{\rho_1} [ i\partial_{\rho_2} ~+~
g \sigma^{a} \CA^{a}_{\rho_2}(x,e) ]\xi(x,e)\Psi(x,e)=0,\nn
\eeqa
where $\CA_{\rho_2}(x,e)=\sigma^{a} \CA^{a}_{\rho_2}(x,e)$.
The Lagrangian density (\ref{generalfermiondensity}) generates the
series of {\it gauge invariant tensor densities
$(\CL^{'}_{\rho_1\rho_2})_{\lambda_1 ... \lambda_{s}}(x)$},
when we expand it in powers of the vector variable $e$:
\be\label{secondseriesdensities}
{{\cal L}}_{\rho_1\rho_2}(x,e) = \sum^{\infty}_{s=0}~
(\CL^{'}_{\rho_1\rho_2})_{\lambda_1 ... \lambda_{s}}(x) ~ e_{\lambda_1}...e_{\lambda_s} .
\ee
The gauge invariant tensor densities
$(\CL^{'}_{\rho_1\rho_2})_{\lambda_1 ... \lambda_{s}}(x)$ allow to construct
two series of gauge invariant Lagrangians:
$ {{\cal L}}_{s+1/2}$ and ${{\cal L}}^{'}_{s+1/2}~~$, s=1,2,.. by the
contraction of the corresponding tensor indices.

The lower gauge invariant tensor density has the form
\beqa\label{secondlowerspinlagrangian}
({{\cal L}}_{\rho_1\rho_2})_{\lambda_1\lambda_2}={1\over 2} \{~&+&
\bar{\psi}_{\lambda_1}   \gamma_{\rho_1} [ i\partial_{\rho_2} ~+~
g A_{\rho_2}  ]\psi_{\lambda_2}   +
\bar{\psi}_{\lambda_2}  \gamma_{\rho_1} [ i\partial_{\rho_2} ~+~
g A_{\rho_2}  ]\psi_{\lambda_1}  + \nonumber\\
&+&\bar{\psi}_{\lambda_1\lambda_2}    \gamma_{\rho_1} [ i\partial_{\rho_2} ~+~
g A_{\rho_2}  ]\psi  +
\bar{\psi}   \gamma_{\rho_1} [ i\partial_{\rho_2} ~+~
g A_{\rho_2}  ]\psi_{\lambda_1\lambda_2}  + \nonumber\\
&+&g \bar{\psi}_{\lambda_1}  \gamma_{\rho_1} A_{\rho_2\lambda_2}  \psi   +
g\bar{\psi}_{\lambda_2}    \gamma_{\rho_1} A_{\rho_2\lambda_1}  \psi   +\nonumber\\
&+&g \bar{\psi}  \gamma_{\rho_1} A_{\rho_2\lambda_2}  \psi_{\lambda_1}   +
g\bar{\psi}   \gamma_{\rho_1} A_{\rho_2\lambda_1} \psi_{\lambda_2}
+ g \bar{\psi}    \gamma_{\rho_1} A_{\rho_2\lambda_1\lambda_2} \psi \},
\eeqa
and we shall use it to generate Lorentz invariant densities. Performing contraction of
the indices of this tensor density with respect to $\eta_{\rho_1\rho_2}\eta_{\lambda_1\lambda_2}$
we shall reproduce our first gauge invariant Lagrangian
density ${\cal L}_{3/2}$ (\ref{firstfermionlagrangianthreehalf}) presented in the previous
section.
We shall get the second gauge invariant Lagrangian performing the
contraction with respect to the $\eta_{\rho_1\lambda_1}\eta_{\rho_2\lambda_2}$,
which is obviously different form the previous one:
\beqa\label{secondfermionlagrangianthreehalf}
{{\cal L}}^{'}_{3/2} &=&{1\over 2}\{
\bar{\psi}_{\mu} \gamma_{\mu} ( i\partial_{\lambda} + g A_{\lambda} )\psi_{\lambda} +
\bar{\psi}_{\lambda} ( i\partial_{\lambda} + g A_{\lambda} )\gamma_{\mu} \psi_{\mu} +\nn\\
&+& \bar{\psi}_{\mu\lambda} \gamma_{\mu} ( i\partial_{\lambda} + g A_{\lambda} )\psi +
\bar{\psi} ( i\partial_{\lambda} + g A_{\lambda} )\gamma_{\mu} \psi_{\mu\lambda} +\\
&+& g \bar{\psi}_{\mu} \gamma_{\lambda}  A_{\mu\lambda} \psi
+g \bar{\psi}  \gamma_{\mu} A_{\lambda\mu} \psi_{\lambda}
+g \bar{\psi}_{\mu} \gamma_{\mu}  A_{\lambda\lambda} \psi
+g \bar{\psi}  \gamma_{\mu} A_{\lambda\lambda} \psi_{\mu}
+g \bar{\psi} \gamma_{\mu}  A_{\lambda\mu\lambda} \psi ~\}.\nn
\eeqa
One can also prove independently from the above consideration,
that these Lagrangian forms are invariant under simultaneous
extended gauge transformations of fermions (\ref{mattertransformation}) and
tensor gauge fields (\ref{polygauge}), calculating their variation:
$$
\delta  \CL^{'}_{3/2} =0.
$$

The currents are given by the variation of the action over the tensor gauge fields:
\beqa\label{secondcurrent}
J^{' a}_{\mu} &=& {1\over 2}g\{\bar{\psi}_{\mu}  \sigma^{a} \gamma_{\lambda}  \psi_{\lambda}+
\bar{\psi}_{\lambda}  \sigma^{a} \gamma_{\lambda}  \psi_{\mu}+
\bar{\psi}_{\mu\lambda}  \sigma^{a} \gamma_{\lambda}  \psi  +
\bar{\psi}  \sigma^{a} \gamma_{\lambda}  \psi_{\mu\lambda}\},\\
J^{' a}_{\mu\nu} &=& {1\over 2}g\{\bar{\psi}_{\mu}  \sigma^{a} \gamma_{\nu}  \psi +
\bar{\psi}  \sigma^{a} \gamma_{\nu}  \psi_{\mu } + \eta_{\mu\nu}
(\bar{\psi}_{\lambda}  \sigma^{a} \gamma_{\lambda}  \psi +
\bar{\psi}  \sigma^{a} \gamma_{\lambda}  \psi_{\lambda} ) \},\nn\\
J^{' a}_{\mu\lambda\rho} &=& {1\over 4}g(
\bar{\psi}  \sigma^{a} \gamma_{\lambda}  \psi ~ \eta_{\mu\rho}+
\bar{\psi}  \sigma^{a} \gamma_{\rho}  \psi ~ \eta_{\mu\lambda}).\nn
\eeqa
From extended gauge invariance it follows that they are divergenceless with respect to the
first indices:
\be
 \partial_{\mu} J^{' a}_{\mu}= \partial_{\mu}J^{' a}_{\mu\nu} =
 \partial_{\mu}J^{' a}_{\mu\lambda\rho} =0.
\ee

Thus the total Lagrangian is a
linear sum of the two Lagrangians: $f_1 {{\cal L}}_{3/2}+ f^{'}_{1}{{\cal L}}^{'}_{3/2} $.
As one can see, from the Lagrangians (\ref{firstfermionlagrangianthreehalf})
and (\ref{secondfermionlagrangianthreehalf}) the interaction of fermions with
tensor gauge bosons is going through the
cubic vertex which includes two fermions and a tensor gauge boson, very
similar to the vertices in QED and the Yang-Mills theory.

\section{\it Euler-Lagrange Equations and Enhanced Symmetry}
As we found above, the total Lagrangian is a
linear sum of the two Lagrangians $f_1 {{\cal L}}_{3/2}+ f^{'}_{1}{{\cal L}}^{'}_{3/2}
= f_1({{\cal L}}_{3/2} +d_1 {{\cal L}}^{'}_{3/2})$
and has the form
\beqa\label{totalfermionlagrangianthreehalf}
{{\cal L}}_{3/2} +d_1 {{\cal L}}^{'}_{3/2} &=&
\bar{\psi}_{\lambda} \gamma_{\mu} ( i\partial_{\mu} + g A_{\mu} )\psi_{\lambda} +
{1\over 2}\bar{\psi} \gamma_{\mu} (i\partial_{\mu} + g A_{\mu} )\psi_{\lambda\lambda}+
{1\over 2}\bar{\psi}_{\lambda\lambda} \gamma_{\mu}
(i\partial_{\mu} + g A_{\mu} )\psi\nonumber\\
&+& g \bar{\psi}_{\lambda} \gamma_{\mu}  A_{\mu\lambda} \psi
+g \bar{\psi}  \gamma_{\mu} A_{\mu\lambda} \psi_{\lambda}
+{1\over 2}g \bar{\psi} \gamma_{\mu}  A_{\mu\lambda\lambda} \psi +\\
 &+&d_1~{1\over 2}\{
\bar{\psi}_{\mu} \gamma_{\mu} ( i\partial_{\lambda} + g A_{\lambda} )\psi_{\lambda} +
\bar{\psi}_{\lambda} ( i\partial_{\lambda} + g A_{\lambda} )\gamma_{\mu} \psi_{\mu} +\nn\\
&+& \bar{\psi}_{\mu\lambda} \gamma_{\mu} ( i\partial_{\lambda} + g A_{\lambda} )\psi +
\bar{\psi} ( i\partial_{\lambda} + g A_{\lambda} )\gamma_{\mu} \psi_{\mu\lambda} +\nn\\
&+& g \bar{\psi}_{\mu} \gamma_{\lambda}  A_{\mu\lambda} \psi
+g \bar{\psi}  \gamma_{\mu} A_{\lambda\mu} \psi_{\lambda}
+g \bar{\psi}_{\mu} \gamma_{\mu}  A_{\lambda\lambda} \psi
+g \bar{\psi}  \gamma_{\mu} A_{\lambda\lambda} \psi_{\mu}
+g \bar{\psi} \gamma_{\mu}  A_{\lambda\mu\lambda} \psi ~\}.\nn
\eeqa
Our aim now is to find out, if there exists a linear combination of these forms
which will produce higher symmetry of the total Lagrangian.
In the weak coupling limit $g \rightarrow 0$ it will take the form
\beqa\label{freefermionlagrangianthreehalf}
{{\cal L}}_{3/2} +d_1 {{\cal L}}^{'}_{3/2} &=&
\bar{\psi}_{\lambda} \gamma_{\mu}  i\partial_{\mu} \psi_{\lambda} +
{1\over 2}\bar{\psi} \gamma_{\mu} i\partial_{\mu} \psi_{\lambda\lambda}+
{1\over 2}\bar{\psi}_{\lambda\lambda} \gamma_{\mu}
i\partial_{\mu}\psi\nonumber\\
 &+&{d_1 \over 2}\{
\bar{\psi}_{\mu} \gamma_{\mu}  i\partial_{\lambda} \psi_{\lambda} +
\bar{\psi}_{\lambda}  i\partial_{\lambda} \gamma_{\mu} \psi_{\mu} +
\bar{\psi}_{\mu\lambda} \gamma_{\mu} i\partial_{\lambda}\psi +
\bar{\psi}  i\partial_{\lambda} \gamma_{\mu} \psi_{\mu\lambda} ~\}.\nn
\eeqa
We have the following free equations of motion:
\beqa\label{freefermionequations}
\gamma_{\mu} i\partial_{\mu} \psi+
{1\over 2}\gamma_{\mu} i\partial_{\mu} \psi_{\lambda\lambda}
+d_1~{1\over 4}(\gamma_{\mu}i\partial_{\lambda} +
\gamma_{\lambda}i\partial_{\mu}) \psi_{\mu\lambda} &=&0 \nn\\
\gamma_{\lambda}  i\partial_{\lambda} \psi_{\mu} +
d_1{1\over 2}  (\gamma_{\mu} i\partial_{\lambda}
+ \gamma_{\lambda}i\partial_{\mu})\psi_{\lambda}&=&0\\
\eta_{\mu\lambda} \gamma_{\rho}  i\partial_{\rho} \psi +
d_1{1\over 2}  (\gamma_{\mu} i\partial_{\lambda}
+ \gamma_{\lambda}i\partial_{\mu})\psi &=&0\nn
\eeqa
or, in equivalent form:
\beqa\label{freefermionequationsshort}
\!\not\!p \psi+
{1\over 2}\!\not\!p \psi_{\lambda\lambda}
+d_1~{1\over 4}(\gamma_{\mu}p_{\lambda} +
\gamma_{\lambda}p_{\mu}) \psi_{\mu\lambda} &=&0 \nn\\
\!\not\!p \psi_{\mu} +
d_1{1\over 2}  (\gamma_{\mu} p_{\lambda}
+ \gamma_{\lambda}p_{\mu})\psi_{\lambda}&=&0\\
\eta_{\mu\lambda} \!\not\!p \psi +
d_1{1\over 2}  (\gamma_{\mu} p_{\lambda}
+ \gamma_{\lambda}p_{\mu})\psi &=&0\nn,
\eeqa
where $\!\not\!p = \gamma_{\mu} p_{\mu}= \gamma_{\mu} i\partial_{\mu} $.
The corresponding total currents $J^{tot}=J + J^{'}$ are equal to the
sum of (\ref{firstcurrent}) and (\ref{secondcurrent}).
Calculating the derivatives of these currents and using equations of motion
one can see, that the conservation of the total currents
over all indices takes place, when $d_1=2$, thus
\beqa
J^{tot~ a}_{\mu}&=& g (\bar{\psi}_{\lambda}  \sigma^{a} \gamma_{\mu}  \psi_{\lambda}+
{1\over 2}\bar{\psi}_{\lambda\lambda}  \sigma^{a} \gamma_{\mu}  \psi  +
{1\over 2}\bar{\psi}  \sigma^{a} \gamma_{\mu}  \psi_{\lambda\lambda}+\nn\\
&&+~\bar{\psi}_{\mu}  \sigma^{a} \gamma_{\lambda}  \psi_{\lambda}+
\bar{\psi}_{\lambda}  \sigma^{a} \gamma_{\lambda}  \psi_{\mu}+
\bar{\psi}_{\mu\lambda}  \sigma^{a} \gamma_{\lambda}  \psi  +
\bar{\psi}  \sigma^{a} \gamma_{\lambda}  \psi_{\mu\lambda}),\nn\\
J^{tot~ a}_{\mu\nu}&=& g ( \bar{\psi}  \sigma^{a} \gamma_{\mu}  \psi_{\nu } +
\bar{\psi}_{\nu}  \sigma^{a} \gamma_{\mu}  \psi +\\
&& +~\bar{\psi}_{\mu}  \sigma^{a} \gamma_{\nu}  \psi +
\bar{\psi}  \sigma^{a} \gamma_{\nu}  \psi_{\mu } + \eta_{\mu\nu}
(\bar{\psi}_{\lambda}  \sigma^{a} \gamma_{\lambda}  \psi +
\bar{\psi}  \sigma^{a} \gamma_{\lambda}  \psi_{\lambda} ) ),\nn\\
J^{tot~ a}_{\mu\lambda\rho}&=& {1\over 2}g(
\bar{\psi}  \sigma^{a} \gamma_{\mu}  \psi ~ \eta_{\lambda\rho}+
\bar{\psi}  \sigma^{a} \gamma_{\lambda}  \psi ~ \eta_{\mu\rho}+
\bar{\psi}  \sigma^{a} \gamma_{\rho}  \psi ~ \eta_{\mu\lambda}),\nn
\eeqa
and we have conservation of the total tensor currents over all indices:
\beqa
\partial_{\nu} J^{tot~ a}_{\nu}&=&0,~~~\nn\\
\partial_{\nu} J^{tot~ a}_{\nu\lambda} &=&0,~~~~
\partial_{\lambda} J^{tot~ a}_{\nu\lambda} =0,\nn\\
\partial_{\nu}  J^{tot~ a}_{\nu\lambda\rho} &=&0,~~~~
\partial_{\lambda}  J^{tot~ a}_{\nu\lambda\rho} =0,~~~~
\partial_{\rho} J^{tot~ a}_{\nu\lambda\rho} =0.
\eeqa
This result is essential for the consistency of the interaction between tensor gauge
bosons and fermions.

It is remarkable, that for a different choice of the coefficient $d_1$ a new type
of gauge symmetry arises.  Let us consider the gauge transformation of the
spinors-tensor fields of the Rarita-Schwinger-Fang-Fronsdal form:
\beqa\label{enhancedfermiontransformation}
\delta \psi  &=& 0 \nn\\
\delta \psi_{\lambda_1} &=& \partial_{\lambda_1} \varepsilon \nn\\
\delta \psi_{\lambda_1 \lambda_2} &=& \partial_{\lambda_1} \varepsilon_{\lambda_2} +
\partial_{\lambda_2} \varepsilon_{\lambda_1}\nn\\
\delta \psi_{\lambda_1 \lambda_2 \lambda_3} &=&
\partial_{\lambda_1} \varepsilon_{\lambda_2\lambda_3} +
\partial_{\lambda_2} \varepsilon_{\lambda_1 \lambda_3} +
\partial_{\lambda_3} \varepsilon_{\lambda_1 \lambda_2}\nn\\
........&=&........................
\eeqa
The variation of the first equation in (\ref{freefermionequationsshort})
will take the form
$$
 \!\not\!p \partial_{\lambda}\varepsilon_{\lambda}
+d_1~{1\over 4}( 2\!\not\!p \partial_{\lambda}\varepsilon_{\lambda} +
2 i \partial^2 \gamma_{\lambda} \varepsilon_{\lambda})
$$
and, if we chose $d_1=-2$ and shall limit the spinor-tensor parameter $\varepsilon_\lambda$ to
fulfil the traceless condition
$$
\gamma_{\lambda} \varepsilon_{\lambda}=0,
$$
the first equation will remain unchanged. As one can clearly see, the rest of the
equations in (\ref{freefermionequationsshort})
are also invariant with respect to the RSFF transformation, if spinor-tensor parameters
$\varepsilon_{\lambda_1 ... \lambda_{s-1}}$ fulfil the traceless conditions:
\be
\gamma_{\lambda} \partial_{\lambda} \varepsilon  =0,~~~~~~~
\gamma_{\lambda_1} \varepsilon_{\lambda_1 \lambda_2 ... \lambda_{s-1}}=0,~~~~s=2,3,...
\ee
In this case the tensor currents will take the form
\beqa
J^{tot~ a}_{\mu}&=& g (\bar{\psi}_{\lambda}  \sigma^{a} \gamma_{\mu}  \psi_{\lambda}+
{1\over 2}\bar{\psi}_{\lambda\lambda}  \sigma^{a} \gamma_{\mu}  \psi  +
{1\over 2}\bar{\psi}  \sigma^{a} \gamma_{\mu}  \psi_{\lambda\lambda}-\nn\\
&&-~\bar{\psi}_{\mu}  \sigma^{a} \gamma_{\lambda}  \psi_{\lambda}-
\bar{\psi}_{\lambda}  \sigma^{a} \gamma_{\lambda}  \psi_{\mu}-
\bar{\psi}_{\mu\lambda}  \sigma^{a} \gamma_{\lambda}  \psi  -
\bar{\psi}  \sigma^{a} \gamma_{\lambda}  \psi_{\mu\lambda}),\nn\\
J^{tot~ a}_{\mu\nu}&=& g ( \bar{\psi}  \sigma^{a} \gamma_{\mu}  \psi_{\nu } +
\bar{\psi}_{\nu}  \sigma^{a} \gamma_{\mu}  \psi -\\
&& -~\bar{\psi}_{\mu}  \sigma^{a} \gamma_{\nu}  \psi -
\bar{\psi}  \sigma^{a} \gamma_{\nu}  \psi_{\mu } - \eta_{\mu\nu}
(\bar{\psi}_{\lambda}  \sigma^{a} \gamma_{\lambda}  \psi +
\bar{\psi}  \sigma^{a} \gamma_{\lambda}  \psi_{\lambda} ) ),\nn\\
J^{tot~ a}_{\mu\lambda\rho}&=& {1\over 2}g(
\bar{\psi}  \sigma^{a} \gamma_{\mu}  \psi ~ \eta_{\lambda\rho}-
\bar{\psi}  \sigma^{a} \gamma_{\lambda}  \psi ~ \eta_{\mu\rho}-
\bar{\psi}  \sigma^{a} \gamma_{\rho}  \psi ~ \eta_{\mu\lambda}).\nn
\eeqa
Corresponding fermion currents are not divergence free, but only traceless part
of the divergence vanishes \cite{rarita,singh1,fronsdal1}.

\section{\it The First Gauge Invariant Lagrangian for Bosons}

We are in a position now to introduce the gauge invariant interaction of
the tensor gauge bosons with the boson  field
$\phi_{\lambda_1 ... \lambda_{s}}(x)$. This set of tensor fields
$\{\phi\}$ contains the scalar
field $\phi$ as one of its family members.
The extended isotopic transformation  of the bosonic matter fields
$\phi_{\lambda_1 ... \lambda_{s}}(x)$
we shall define by the formulas
\cite{Savvidy:2005fi,Savvidy:2005zm,Savvidy:2005ki}
\beqa\label{scalartransformation}
\delta_{\xi}  \phi  &=&-i \tau^{a}\xi^{a}  \phi ,\nonumber\\
\delta_{\xi}  \phi_{\lambda} &=& -i \tau^{a}( \xi^{a} ~\phi_{\lambda}   +
\xi^{a}_{\lambda}~ \phi) ,\nonumber\\
\delta_{\xi}  \phi_{\lambda\rho} &=& -i \tau^{a}( \xi^{a} ~\phi_{\lambda\rho}   +
\xi^{a}_{\lambda}~ \phi_{\rho} + \xi^{a}_{\rho}~ \phi_{\lambda} +
\xi^{a}_{\lambda\rho}~ \phi ),\\
........&.&.......................,\nn
\eeqa
where $\tau^{a}$ are the matrices of the representation $\tau$ of the
compact Lie group G, according to which the whole family of $\phi's$ transforms.
There is an essential difference
in the transformation properties of the tensor gauge fields
$A_{\mu\lambda_1 ... \lambda_{s}}$
versus
$\phi_{\lambda_1 ... \lambda_{s}}$ . The transformation law for the
bosonic matter fields (\ref{scalartransformation}) is homogeneous,
whereas the transformation of the tensor gauge  fields (\ref{polygauge})
is inhomogeneous. The general form of the
above transformation is:
\beqa
\delta_{\xi}  \phi_{\lambda_1 ... \lambda_{s}}(x) &=& -i \sum^{s}_{i=0}  \sum_{P's}
 \xi_{\lambda_1 ... \lambda_{i}} ~\phi_{\lambda_{i+1} ... \lambda_{s}}(x), ~~~~~~~
 s=0,1,2,...~~~,
\eeqa
and the invariant quadratic form is:
\be\label{invariantform}
U(\phi) = \sum^{\infty}_{s=0} \lambda_{s+1} U_s(\phi),~~~~~~~U_s(\phi) = \sum^{2s}_{i=0}~a^{s}_i ~
\phi^{+}_{\lambda_1 ... \lambda_i}~
\phi_{\lambda_{i+1}...\lambda_{2s}}
\sum_{p's} \eta^{\lambda_{i_1} \lambda_{i_2}} .......
\eta^{\lambda_{i_{2s-1}} \lambda_{i_{2s}}}~,
\ee
where $\lambda_s$  are arbitrary coupling constants and the sum $\sum_p$ runs
over all permutations of $p's$ and
the numerical coefficient $a^{s}_i = s!/i!(2s-i)!$,~ $\lambda_1 =1$. Notice that
the number of real gauge parameters $\xi^{a}_{\lambda_1 ... \lambda_{s}}$
is proportional to the dimension
dimG of the compact Lie group G, while the number of tensor matter fields
$\phi^{i}_{\lambda_1 ... \lambda_{s}}$ is proportional to the dimension of the
representation $\tau^{a}_{ij}$ of the group G.  Because they are totally symmetric tensors,
they have the same space-time dimensions, thus
$$
dim \xi = dimG  \times ~dimT, ~~~~dim\phi = dim\tau \times ~dimT,
$$
where $dimT$ is the dimension of the totally symmetric rank-s tensor.

The invariant Lagrangian for scalar field is
$$
{{\cal L}}^{B}_{0} =   -\nabla^{ij}_{\mu}\phi^{+j} ~ \nabla^{ik}_{\mu}\phi^{k} ~-U(\phi),
$$
where $\nabla^{ij}_{\mu} = \delta^{ij} \partial_{\mu} -ig \tau^{ij}_{a} A^{a}_{\mu}$
and for the rank-one field it
has the form
\cite{Savvidy:2005fi,Savvidy:2005zm,Savvidy:2005ki}:
\beqa\label{scalarlagrangian}
-\CL^{B}_{1} &=&
\nabla_{\mu}\phi^{+}_{\lambda} ~\nabla_{\mu}\phi_{\lambda} +
{1\over 2}\nabla_{\mu}\phi^{+}_{\lambda\lambda} ~ \nabla_{\mu}\phi +
{1\over 2}\nabla_{\mu}\phi^{+} ~ \nabla_{\mu}\phi_{\lambda\lambda} +\nonumber\\
&-& ig \nabla_{\mu}\phi^{+} ~A_{\mu\lambda}\phi_{\lambda} +
ig \phi^{+}_{\lambda} A_{\mu\lambda}~\nabla_{\mu}\phi -
ig \nabla_{\mu}\phi^{+}_{\lambda} ~A_{\mu\lambda}\phi+
ig \phi^{+} A_{\mu\lambda}  ~ \nabla_{\mu}\phi_{\lambda}+\nn\\
&+&g^2  \phi^{+} A_{\mu\lambda} A_{\mu\lambda}\phi-
{1\over 2}ig \nabla_{\mu}\phi^{+} ~A_{\mu\lambda\lambda}\phi +
{1\over 2}ig \phi^{+} A_{\mu\lambda\lambda}~\nabla_{\mu}\phi ~+~U(\phi).
\eeqa
The variation of the Lagrangian is equal to zero, $\delta \CL^{B}_{1} =0$.
In the next section we shall see, that there exists a second invariant form
 which can be constructed in terms of boson fields
and the total Lagrangian is a linear sum.

\section{\it The Second Gauge Invariant Lagrangian for Bosons }

The Lagrangian (\ref{scalarlagrangian}) is not the
most general Lagrangian which can be constructed in terms
of the above boson fields $\phi_{\lambda_1 ... \lambda_{s}}(x)$.
As we shall see, here also exists a second invariant form $\CL^{'B}_{1}$
which can be constructed in terms of boson
fields $\phi_{\lambda_1 ... \lambda_{s}}(x)$  and the total Lagrangian is a
linear sum of them:  $ b_1 \CL^{B}_{1} + b^{'}_1 \CL^{'B}_{1} $.
The sum exhibits additional gauge invariance
with specially chosen coefficient $ b^{'}_1 $ .

Let us consider the gauge invariant tensor density of the form
\cite{Savvidy:2005fi,Savvidy:2005zm,Savvidy:2005ki}
\be\label{secondseriesbosons}
{{\cal L}}_{\rho_1\rho_2} =     \nabla^{ij}_{\rho_1}(e)\Phi^{+j}(x,e) ~
\nabla^{ik}_{\rho_2}(e) \Phi^{k}(x,e) ,
\ee
where $\nabla^{ij}_{\mu}(e) = \delta^{ij} \partial_{\mu} -
ig \tau^{ij}_{a} \CA^{a}_{\mu}(x,e)$.
The Lagrangian density (\ref{secondseriesbosons}) generates the
second series of {\it gauge invariant tensor densities
$(\CL_{\rho_1\rho_2})_{\lambda_1 ... \lambda_{s}}(x)$},
when we expand it in powers of the vector variable $e_\lambda$:
\be\label{secondseriesdensities}
{{\cal L}}_{\rho_1\rho_2}(x,e) = \sum^{\infty}_{s=0}~
(\CL_{\rho_1\rho_2})_{\lambda_1 ... \lambda_{s}}(x) ~ e_{\lambda_1}...e_{\lambda_s} .
\ee
The lower gauge invariant tensor density has the form
\beqa\label{secondlowerspinlagrangian}
&&({{\cal L}}_{\rho_1\rho_2})_{\lambda_1\lambda_2}=\nn\\
&+&{1\over 2} \{~
\nabla_{\rho_1}  \phi^{+}_{\lambda_1}  \nabla_{\rho_2}  \phi_{\lambda_2}  +
\nabla_{\rho_1}  \phi^{+}_{\lambda_2}  \nabla_{\rho_2}  \phi_{\lambda_1}  +
\nabla_{\rho_1}  \phi^{+}_{\lambda_1\lambda_2}  \nabla_{\rho_2}  \phi  +
\nabla_{\rho_1}  \phi^{+}  \nabla_{\rho_2}  \phi_{\lambda_1\lambda_2}  \\
&-& i g \nabla_{\rho_1}  \phi^{+}_{\lambda_1}  A_{\rho_2\lambda_2}  \phi
-i g \nabla_{\rho_1}  \phi^{+}_{\lambda_2}  A_{\rho_2\lambda_1}  \phi
-i g \nabla_{\rho_1}  \phi^{+}  A_{\rho_2\lambda_1}  \phi_{\lambda_2}
-i g \nabla_{\rho_1}  \phi^{+}  A_{\rho_2\lambda_2}  \phi_{\lambda_1}
\nonumber\\
&+& i g \phi^{+}_{\lambda_1} A_{\rho_1\lambda_2}\nabla_{\rho_2}  \phi
+i g \phi^{+}_{\lambda_2} A_{\rho_1\lambda_1}\nabla_{\rho_2}     \phi
+i g \phi^{+}   A_{\rho_1\lambda_1}  \nabla_{\rho_2}    \phi_{\lambda_2}
+i g  \phi^{+}  A_{\rho_1\lambda_2} \nabla_{\rho_2}  \phi_{\lambda_1}
\nonumber\\
&+&g^2 \phi^{+}  A_{\rho_1\lambda_1} A_{\rho_2 \lambda_2}  \phi
+ g^2 \phi^{+}  A_{\rho_1\lambda_2} A_{\rho_2 \lambda_1 }  \phi
-i g \nabla_{\rho_1}  \phi^{+}  A_{\rho_2\lambda_1\lambda_2}  \phi
+i g \phi^{+} A_{\rho_1\lambda_1\lambda_2} \nabla_{\rho_2}    \phi
\}, \nn
\eeqa
and by an appropriate contraction of indices generates Lorentz invariant densities.
Performing contraction of
the indices of this tensor density with respect to $\eta_{\rho_1\rho_2}\eta_{\lambda_1\lambda_2}$
we shall get our first gauge invariant Lagrangian
density $\CL^{B}_{1}$ (\ref{scalarlagrangian}) presented in the previous
section.

We shall get the second gauge invariant Lagrangian performing the
contraction with respect to $\eta_{\rho_1\lambda_1}\eta_{\rho_2\lambda_2}$,
which is obviously different from the previous one:
\beqa\label{secondlowerspinlagrangianscalar}
\CL^{'B}_{1}&=&{1\over 2} \{~
\nabla_{\mu}  \phi^{+}_{\mu}  \nabla_{\lambda}  \phi_{\lambda}  +
\nabla_{\mu}  \phi^{+}_{\lambda}  \nabla_{\lambda}  \phi_{\mu}  +
\nabla_{\mu}  \phi^{+}_{\mu\lambda}  \nabla_{\lambda}  \phi  +
\nabla_{\mu}  \phi^{+}  \nabla_{\lambda}  \phi_{\mu\lambda}  -\\
&-& i g \nabla_{\mu}  \phi^{+}_{\mu}  A_{\lambda\lambda}  \phi
-i g \nabla_{\mu}  \phi^{+}_{\lambda}  A_{\lambda\mu}  \phi
-i g \nabla_{\mu}  \phi^{+}  A_{\lambda\mu}  \phi_{\lambda}
-i g \nabla_{\mu}  \phi^{+}  A_{\lambda\lambda}  \phi_{\mu}+
\nonumber\\
&+& i g \phi^{+}_{\mu} A_{\lambda\lambda}\nabla_{\mu}  \phi
+i g \phi^{+}_{\mu} A_{\mu\lambda}\nabla_{\lambda}     \phi
+i g \phi^{+}   A_{\mu\mu}  \nabla_{\lambda}    \phi_{\lambda}
+i g  \phi^{+}  A_{\mu\lambda} \nabla_{\lambda}  \phi_{\mu}+
\nonumber\\
&+&g^2 \phi^{+}  A_{\mu\mu} A_{\lambda\lambda}  \phi
+ g^2 \phi^{+}  A_{\mu\lambda} A_{\lambda\mu }  \phi
-i g \nabla_{\mu}  \phi^{+}  A_{\lambda\mu\lambda}  \phi
+i g \phi^{+} A_{\lambda\mu\lambda} \nabla_{\mu}    \phi
\}. \nn
\eeqa
Thus the total Lagrangian $ {{\cal L}}^B_{0}+b(\CL^{B}_{1}+h \CL^{'B}_{1})$
has a linear sum of two forms and our aim is to find out, if there exists a linear combination of these forms
which admits additional higher symmetries.
In zero coupling limit it will take the form
\beqa\label{freebosonlagrangian}
 \CL^{B}_{1}+h \CL^{'B}_{1}=
&-&\partial_{\mu}\phi^{+}_{\lambda} ~\partial_{\mu}\phi_{\lambda}
+{h\over 2} \{~\partial_{\mu}  \phi^{+}_{\mu}  \partial_{\lambda}  \phi_{\lambda}  +
\partial_{\mu}  \phi^{+}_{\lambda}  \partial_{\lambda}  \phi_{\mu} \} -\nn\\
&-&{1\over 2}\partial_{\mu}\phi^{+}_{\lambda\lambda} ~ \partial_{\mu}\phi -
{1\over 2}\partial_{\mu}\phi^{+} ~ \partial_{\mu}\phi_{\lambda\lambda}
+{h\over 2} \{~
\partial_{\mu}  \phi^{+}_{\mu\lambda}  \partial_{\lambda}  \phi  +
\partial_{\mu}  \phi^{+}  \partial_{\lambda}  \phi_{\mu\lambda} \} .\nn
\eeqa
If we take $h=1$, it will reduce to the form
\beqa\label{freebosonlagrangianlast}
\CL^{B}_{1}+ \CL^{'B}_{1}=
&-&\partial_{\mu}\phi^{+}_{\lambda} ~\partial_{\mu}\phi_{\lambda}
+\partial_{\mu}  \phi^{+}_{\lambda}  \partial_{\lambda}  \phi_{\mu}  -\nn\\
&-&{1\over 2}\partial_{\mu}\phi^{+}_{\lambda\lambda} ~ \partial_{\mu}\phi -
{1\over 2}\partial_{\mu}\phi^{+} ~ \partial_{\mu}\phi_{\lambda\lambda}
+{1\over 2} \partial_{\mu}  \phi^{+}_{\mu\lambda}  \partial_{\lambda}  \phi  +
{1\over 2} \partial_{\mu}  \phi^{+}  \partial_{\lambda}  \phi_{\mu\lambda}
\eeqa
and  become invariant with respect to the gauge transformation
of the form
\beqa\label{additionalgaugesymmetry}
\phi  &\rightarrow & \phi  \nn\\
\phi_{\mu} &\rightarrow &\phi_{\mu} + \partial_{\mu} \omega \nn\\
\phi_{\mu\nu} &\rightarrow &\phi_{\mu\nu} + \partial_{\mu} \omega_{\nu} +
\partial_{\nu} \omega_{\mu}.
\eeqa
This symmetry transformation is an enhanced symmetry of the Lagrangian. The
original system was invariant under the gauge transformation
(\ref{scalartransformation}).

This  phenomenon of enhancement of the original symmetries
is of the same nature as we have already observed
in the case of tensor gauge fields and fermions, where the extended
gauge transformation (\ref{polygauge}) and (\ref{mattertransformation}) have been
enhanced to larger
symmetries (\ref{fullgroupofextendedtransformation}) and
(\ref{enhancedfermiontransformation}).
The above enhanced gauge symmetries
allow to exclude negative norm states from our system of tensor fields
and in the given case the zero component of the boson field $\phi_{\mu}$.

\section{\it Symmetry Breaking and Masses of Tensor Gauge Bosons}

The  Lagrangian
${{\cal L}}= {{\cal L}}^B_{0}+ \CL^{B}_{1}+\CL^{'B}_{1}$
can be responsible for the mass generation of the second-rank tensor gauge field
$A^{a}_{\mu\nu}$. The relevant terms have the following form:
\beqa
\CL = &-&\nabla_{\mu}\phi^{+} ~ \nabla_{\mu}\phi - U(\phi)\nn\\
&+& b_1  \{ - \nabla_{\mu}\phi^{+}_{\lambda} ~\nabla_{\mu}\phi_{\lambda}
+{1\over 2}\nabla_{\mu}  \phi^{+}_{\mu}  \nabla_{\lambda}  \phi_{\lambda}
+{1\over 2}\nabla_{\mu}  \phi^{+}_{\lambda}  \nabla_{\lambda}  \phi_{\mu}\nn\\
&-&g^2  \phi^{+} (A_{\mu\lambda} A_{\mu\lambda}
-{1\over 2}   A_{\mu\mu} A_{\lambda\lambda}
- {1\over 2} A_{\mu\lambda} A_{\lambda\mu } )  \phi \}.
\eeqa
The first term describes the standard interaction of the charge vector gauge boson
with charged scalar field and with properly chosen scalar potential will
generate the mass of the vector boson (see the end of this section).
The next three terms describe the
interaction of the charged vector gauge bosons $A_{\mu}$ with charged
vector boson $\phi_{\lambda}$ and the last three terms describe the
interaction of the charged tensor gauge bosons $A_{\mu\nu}$ with charged
scalar boson $\phi$. When the scalar field gets the vacuum expectation
value (\ref{generalizedpotensioal}), the charged tensor gauge bosons receive the mass term of the form
\beqa\label{masstermsforgaugebososns}
b_1  g^2  \phi^{+} (A_{\mu\lambda} A_{\mu\lambda}
-{1\over 2}   A_{\mu\mu} A_{\lambda\lambda}
- {1\over 2} A_{\mu\lambda} A_{\lambda\mu } )\phi \rightarrow \nn\\
b_1  g^2 <\phi^{\dag}\tau_{a}> <\tau_{b}\phi> (A^{a}_{\mu\lambda} A^{b}_{\mu\lambda}
-{1\over 2}   A^{a}_{\mu\mu} A^{b}_{\lambda\lambda}
- {1\over 2} A^{a}_{\mu\lambda} A^{b}_{\lambda\mu } ).
\eeqa
Decomposing the tensor gauge field $A_{\mu\nu}$ into symmetric and
antisymmetric parts $A_{\mu\nu}= {1\over 2}(A_{\mu\nu} +A_{\nu\mu})
+{1\over 2}(A_{\mu\nu} -A_{\nu\mu})= T_{\mu\nu} +S_{\mu\nu}$
one can see that the mass term takes the form
\beqa
{m^{2}_{T}\over 2} (T_{\mu\lambda} T_{\mu\lambda}
-  T_{\mu\mu} T_{\lambda\lambda})
+ {m^{2}_{S} \over 2}    S_{\mu\lambda} S_{\lambda\mu },
\eeqa
where the mass matrices are
\be
m^{2}_{T} =   ({b_1 \over g_2})  g^2  <\phi^{\dag}\tau_{a}> <\tau_{b}\phi>,~~~~~
m^{2}_{S}= 3 ({b_1 \over g_2}) g^2  <\phi^{\dag}\tau_{a}> <\tau_{b}\phi>,
\ee
and we conclude that the coupling constant $b_1 $ should be positive.
As we can see from above formulas, the symmetric $T_{\mu\nu}$ and antisymmetric $S_{\mu\nu}$
parts of the tensor gauge field
get different masses: the antisymmetric part gets the mass which is three
times bigger than that of the symmetric tensor gauge boson.
The coupling constant $b_1 $, as we discussed earlier, remains arbitrary in this
model, therefore the relation between masses of the tensor gauge bosons and
vector gauge bosons $m^{2}_{V}= 2 g^2  <\phi^{\dag}\tau_{a}> <\tau_{b}\phi>$
is given by the relations
\be\label{tensotmassretio}
m^{2}_{T} = ({b_1 \over 2 g_2}) m^{2}_{V},~~~~
m^{2}_{S} = 3 ({b_1 \over 2 g_2}) m^{2}_{V},
\ee
with the {\it $b_1 $-independent} mass ratio
\be
{m^{2}_{S} \over m^{2}_{T} } = 3,
\ee
which is an interesting prediction
of this model\footnote{These mass formulas are written in preposition that $g_2 \neq 1$
in (\ref{fulllagrangian2}), that is, the kinetic term of the tensor gauge field
$A_{\mu\nu}$   is normalized to $-(1/4)g_2$.}.

We have to introduce the invariant self-interaction Lagrangian for the extended scalar
sector. The first two quadratic forms, which are invariant with respect to the extended
homogeneous transformations (\ref{scalartransformation}), have the form
(\ref{invariantform})
\be\label{invariantquadraticform}
U(\phi)= \phi^{\dag} \phi +
\lambda_2 ( \phi^{\dag}_{\mu}\phi_{\mu} + {1\over 2}
\phi^{\dag}\phi_{\mu\mu}  +{1\over 2}
\phi^{\dag}_{\mu\mu}  \phi ).
\ee
Its invariance can be confirmed by direct calculation similar to the one we
performed above. Using this quadratic form we can construct the invariant
potential as
\be\label{generalizedpotensioal}
U(\phi) ={1\over 4} \lambda^2 [\phi^{\dag} \phi -   \eta^2]^2 ~+
{1\over 4} \lambda^2 [\phi^{\dag} \phi +\lambda_2 (\phi^{\dag}_{\lambda}\phi_{\lambda} + {1\over 2}
\phi^{\dag}\phi_{\lambda\lambda}  +{1\over 2}
\phi^{\dag}_{\lambda\lambda}  \phi )]^2
\ee
so that the vacuum expectation value of the scalar field will be as in the
standard model:
$$
<\phi>_{vac} = \eta /\sqrt{2}.
$$
The Higgs boson mass therefore
remains the same as in the standard model:
$$
m_H = \lambda \eta.
$$
The  vector  boson $\phi_{\lambda}$ can also acquire mass through the interaction term:
\be
{1\over 4} ~\lambda^2 ~2~\lambda_2 ~ \phi^{\dag} \phi ~\phi^{\dag}_{\mu}\phi_{\mu}~~~
\rightarrow ~~~{1\over 2} \lambda^2 \lambda_2 <\phi^{\dag}>
<\phi> \phi^{\dag}_{\mu}\phi_{\mu}=
\lambda_2 {\lambda^2 \eta^2 \over 4} \phi^{\dag}_{\mu}\phi_{\mu},
\ee
and it is proportional to the mass of the standard Higgs scalar:
\be\label{higgsvectorbosonmass}
m^2_{\phi} = ({\lambda_2 \over 4 b_1 }) ~ m^2_H.
\ee
We see that $\lambda_2$ should be positive.
This formula is of the same nature as for the tensor gauge bosons (\ref{tensotmassretio})
and reflects the fact that masses of higher-spin partners can be expressed through masses
of the standard model particles and the coupling constants between them. In the given case
these coupling constants are $b_1 $  (\ref{scalarlagrangian}),(\ref{secondlowerspinlagrangianscalar})
and $\lambda_2$ (\ref{invariantform}),(\ref{generalizedpotensioal}).

This work was partially supported by ENRAGE (European Network on Random
Geometry), a Marie Curie Research Training Network, contract MRTN-CT-2004-
005616.

\vfill
\end{document}